\documentclass[prb,showpacs,twocolumn]{revtex4}
\newcommand {\TiO}{Ti$_2$O$_3$}  
\newcommand {\TiVO}{(Ti$_{1-x}$V$_x$)$_2$O$_3$}  

\newcommand {\ttwog}{$t_{2g}$}

\newcommand {\HparaC}{$H\parallel c$}
\newcommand {\HperpC}{$H\perp c$}

\newcommand {\Tw}{$\theta_{\mathrm{w}}$}
\newcommand {\Tg}{$T_{\mathrm{g}}$}
\usepackage{graphicx}
\usepackage{bm}
\usepackage{pifont}
\usepackage{amsmath}
\begin{document}
\title{Large Magnetoresistance and Spin-Polarized Heavy-Mass Electron State in a Doped Valence Bond Solid {\TiVO}}
\author{Masaki Uchida$^1$, Yoshinori Onose$^{1,2}$, and Yoshinori Tokura$^{1,2,3}$} 
\affiliation{$^1$Department of Applied Physics, University of Tokyo, Tokyo 113-8656, Japan \\ $^2$Multiferroics Project, ERATO, Japan Science and Technology Agency (JST), Tokyo 113-8656, Japan \\ $^3$Cross-Correlated Materials Research Group (CMRG) and Correlated Electron Research Group (CERG), ASI, RIKEN, Wako 351-0198, Japan}
\date{\today}

\begin{abstract}
A heavy-mass electron state is realized in a doped valence bond solid {\TiVO}.
In this system, itinerant holes mediate the mainly ferromagnetic RKKY interaction between the localized magnetic moments and
become readily spin-polarized under a magnetic field,
while showing large negative magnetoresistance.
In spite of the ferromagnetic interaction among the carriers,
their effective mass is found to be 1 or 2 orders of magnitude larger than that of usual doped semiconductors.
Such strong mass renormalization is ascribable to the polaron formation on the Ti-dimer,
where the spin-singlet state is originally formed.
Doping dependence of the electronic specific-heat coefficient implies that the dimeric lattice fluctuation or softening is responsible for the enhanced electron-phonon interaction.
\end{abstract}
\pacs{71.27.+a, 71.38.-k, 75.47.-m, 75.50.Lk}
\maketitle

\begin{figure}
\begin{center}
\includegraphics*[width=8.5cm]{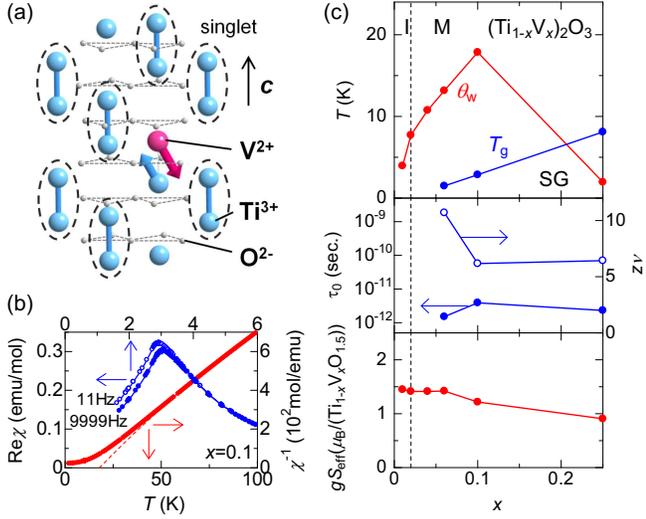}
\caption{(Color online).
(a) Crystal lattice with schematic spin structure in {\TiVO}.
Ti and V ions are nominally trivalent ($S=1/2$) and divalent ($S=3/2$), respectively.
Ti ions in the dimer form a singlet pair along the $c$-axis.
(b) In-phase ac magnetic susceptibility measured at 11 and 9999 Hz by 0.3 mT and
inverse susceptibility measured by 1 T for {\HperpC} for {\TiVO} ($x=0.1$).
A dashed curve indicates the fitting by the Curie-Weiss law with a $T$-independent term.
(c) Doping variation of the Weiss temperature {\Tw}, the spin-glass (SG) transition temperature {\Tg} (upper panel),
the spin flipping time $\tau _0$, the product of critical exponents $z\nu$ (middle panel),
and the effective magnetic moment $gS_{\mathrm{eff}}$ (lower panel) deduced from the Curie-Weiss plot for {\TiVO}.
A vertical dashed line represents the insulator (I)-metal (M) phase boundary.
}
\label{fig1}
\end{center}
\end{figure}

Strong electron correlation in a solid with an integer electron (hole) number per atomic site causes a Mott insulator state,
in which charge, spin, and orbital degrees of freedom are all active and play important roles in the possible Mott transition. \cite{MIT}
Doping carriers into the Mott insulator can produce the barely metallic state with emergent properties
such as high-temperature superconductivity \cite{super,RVB} and colossal magnetoresistance. \cite{CMR}
The conventional Mott insulator exhibits antiferromagnetic or ferromagnetic long-range order,
while a nonmagnetic state based on the spin-singlet formation sometimes shows up in a specific lattice form and/or by magnetic frustration.
This quantum state, termed a valence bond solid (VBS), \cite{VBS}
can also provide fertile ground for studying intriguing phenomena.
For example, the carrier doping effect on the VBS system has long attracted much attention,
especially in light of carrier dynamics as in the resonant valence bond solid state;
the possible superconductivity has also been of interest. \cite{RVB}
The purpose of this study is to investigate the dynamics of charge carriers doped into {\TiO},
a classic example of the VBS, in which the local spin singlets are formed on dimeric Ti sites.

In Fig. 1(a), we show the crystal lattice with schematic spin structure in {\TiVO}.
{\TiO} has an $\alpha$-corundum structure, in which the nearest neighbor Ti ions form pairs along the $c$ axis. 
As a result, two $S=1/2$ spins on the Ti-Ti dimer form a singlet state at the ground state.
The dimerized state is viewed as the VBS;
the parent compound {\TiO} is thus insulating.
Substituting Ti with V induces the insulator-metal transition \cite{resV} as a consequence of change in band filling (hole doping).
The carriers were confirmed to be holes by the sign of the Hall coefficient. \cite{hall_Eg}
The singlet pairs are broken into V-doped dimers,
where the $S=1/2$ and $3/2$ local moments are revived at the Ti and V sites, respectively.
These moments are antiferromagnetically coupled with each other and behave as a $S=1$ local moment, as noted later.
The Ruderman-Kittel-Kasuya-Yosida (RKKY) interaction mediated by the itinerant holes works on the local moments
to induce the canonical spin-glass phase in the metallic region of {\TiVO}. \cite{TVOsg1,TVOsg2,TVOsg3}
In this paper, we report that the doped VBS {\TiVO} shows a large negative magnetoresistance and a heavy-mass electron state
likely caused by the polaronic coupling to the dimeric distortion.
The ferromagnetic exchange interaction among the localized spins is mediated by the conduction holes with
low carrier density (or small Fermi wavenumber) and heavy mass.
Thus, the present result shows that {\TiVO} can be viewed as a magnetic semiconductor based on the VBS.

Single crystals of {\TiVO} were grown with a floating-zone method in a forming gas flow of Ar (93$\%$) and H$_2$ (7$\%$).
The resistivity, magnetization, heat capacity, and Hall coefficient were measured with the Quantum Design Physical Property Measurement System.

In Fig. 1(b), we illustrate the in-phase ac magnetic susceptibility and inverse susceptibility for the doping concentration of $x=0.1$.
The ac magnetic susceptibility shows clear frequency dependence and we obtain the spin-glass transition temperature {\Tg} = 2.9 K in the low frequency limit
by applying the dynamical scaling relation $\tau /\tau _0 = ((T_{\mathrm{f}}-T_{\mathrm{g}})/T_{\mathrm{g}})^{-z\nu}$
in the frequency region of 11-9999 Hz.
Here, $\tau$, $\tau _0$, $T_{\mathrm{f}}$, and $z\nu$ denote
the observation time (inverse of the ac frequency), the spin flipping time, the peak temperature measured at each frequency,
and the product of critical exponents, respectively.
On the other hand, the inverse susceptibility curve at high temperature is well fitted by the Curie-Weiss law,
$\chi =C/(T-\theta_{\mathrm{w}}) +\chi _0$,
where $C$, {\Tw}, and $\chi _0$ are the Curie constant, the Weiss temperature, and the $T$-independent constant term including Pauli and Van Vleck paramagnetic terms, respectively.
Based on these analyses, 
we show in Fig. 1(c) the doping variation of the magnetic properties in the metallic phase of {\TiVO}.
The canonical spin glass phase induced by V-doping appears over a wide doping region up to $x\sim 0.4$. \cite{TVOsg1}
Especially below $x\sim 0.25$, the sign of {\Tw} is positive reflecting the dominant ferromagnetic interaction between the local moments.
The spin flipping time $\tau _0$ is $\sim 10^{-12}$ s, indicating that an atomic-scale spin glass state is realized in this system.
According to the Monte Carlo calculation results, \cite{Ising,Heisenberg}
the change of $z\nu$ can be interpreted as crossover from anisotropic Ising-like ($z\nu \sim 10.5$) to Heisenberg-like ($z\nu \sim 5.5$) behavior
with increasing V-doping level.
The effective magnetic moment $gS_{\mathrm{eff}}$ $(\simeq 1.4  \mu _{\mathrm{B}})$ also defined by the Curie-Weiss fit indicates
that the spins of the Ti$^{3+}$ ($S=1/2$) and V$^{2+}$ ($S=3/2$) ions on the dimer couple antiferromagnetically
to form $S=1$ localized spin up to high temperature.
The small discrepancy between the expected effective moment ($gS_{\mathrm{eff}} \simeq 2 \mu _{\mathrm{B}}$)
and the observed value may be caused by the valence fluctuation between V$^{2+}$ and V$^{3+}$
concomitant with the valence change (electron itinerancy) at the Ti site on the dimer.


\begin{figure}
\begin{center}
\includegraphics*[width=8.5cm]{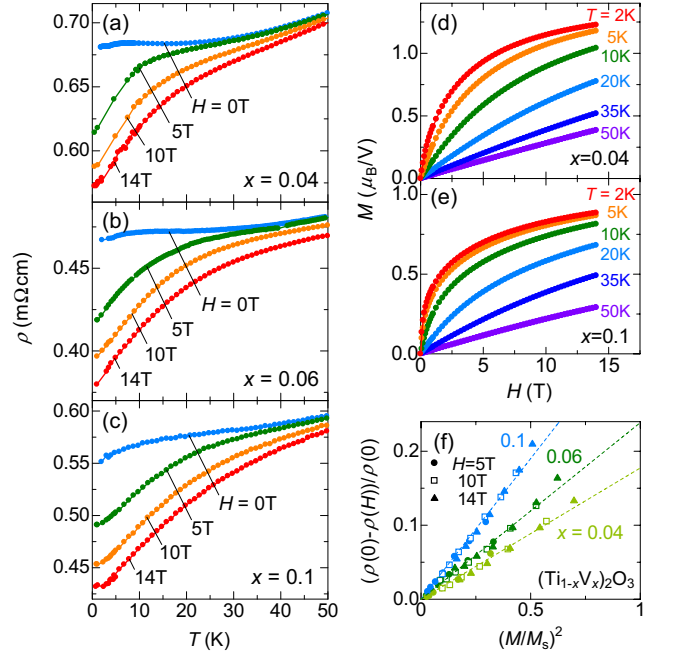}
\caption{(Color online).
(a)-(c) Magnetoresistance at various applied magnetic fields $H$ in the $I$ (current) $\parallel H \parallel c$ configuration for {\TiVO} ($x=0.04, 0.06,$ and 0.1).
(d)-(e) $M-H$ curves at {\HparaC} for {\TiVO} ($x=0.04$ and 0.1).
(f) Magnitude of negative magnetoresistance $(\rho (0)-\rho (H))/\rho (0)$ as a function of $(M/M_{\mathrm{s}})^2$.
$M_{\mathrm{s}}$ is the saturated magnetization, which was estimated by the Curie-Weiss fitting for high-temperature susceptibility.
Filled circles, open squares, and filled triangles represent the data at $H=5, 10,$ and 14 T, respectively.
}
\label{fig2}
\end{center}
\end{figure}
 
In Figs. 2(a)-(c), we show the magnetoresistance for {\TiVO} ($x=0.04, 0.06,$ and 0.1) at various magnetic fields.
A large negative magnetoresistance appears below around 50 K and its magnitude increases with lowering temperature and increasing magnetic field.
With increasing the doping level, the magnitude of the magnetoresistance ratio increases
and reaches about 20 \% at the lowest temperature and highest magnetic field (14 T) for $x=0.1$.
Such a large negative magnetoresistance is rare among transition-metal oxides only with {\ttwog} electrons.

As shown in Figs. 2(d) and (e), each magnetization curve at $T=2$ K shows asymptotical approach to a maximum value like a hard ferromagnet
and the value almost saturates at $H=14$ T.
The nearly saturated value is in accord with the effective magnetic moment $gS_{\mathrm{eff}}$ obtained by fitting the high-temperature susceptibility with the Curie-Weiss law.
In Fig. 2(f), we plot the magnitude of the negative magnetoresistance $(\rho (0)-\rho (H))/\rho (0)$ as a function of $(M/M_{\mathrm{s}})^2$ for {\TiVO}.
Here, $M_{\mathrm{s}}$ is the hypothetical saturated magnetization corresponding to $gS_{\mathrm{eff}}$.
In various magnetoresistive compounds including colossal magnetoresistive manganites,
its magnitude is observed to obey the scaling function $(\rho (0)-\rho (H))/\rho (0)=C(M/M_{\mathrm{s}})^2$, \cite{Msmanganite}
where $C$ represents a coupling strength between the itinerant holes and the localized spins. \cite{Msdouble,Mssd,Mskondo}
Also in {\TiVO}, the negative magnetoresistance shows an $M^2$-dependence
and the data at different temperatures merge into universal ones.
This indicates that the itinerant holes should be highly spin-polarized via the coupling with local spins under a high enough magnetic field.

In Fig. 3(a), we show the specific heat at $H=14$ T in the form of a $C/T$ vs $T^2$ plot.
The magnetic specific heat in the spin-glass state has a $T$-linear dependence below {\Tg},
but is suppressed by applying the high magnetic field.
We confirmed that the  $C/T-T^2$ slope near the lowest temperature is close to the value calculated from the Debye temperature of {\TiO}. \cite{debye}
The obtained electronic specific-heat coefficient $\gamma$ (upper panel of Fig. 3(d)) is
1 order of magnitude larger than those of conventional metals.
As an example, we show the temperature variation of the Hall resistivity for {\TiVO} with $x=0.02$ and $x=0.1$ in Fig. 3(b).
The linear field dependence of the Hall resistivity suggests the negligible anomalous Hall effect
and ensures the estimation of the carrier density from the slope.
As shown in Fig. 3(c), the obtained carrier density $n$ per one metal (M)-site in {\TiVO} is very small
and asymptotically approaches $\sim x$ at high temperatures.
This is in sharp contrast to the canonical Mott transition systems,
where $n$ becomes close to $1-x$, rather than $x$, upon the Mott transition. \cite{TiMIT}
The weight of the Drude component estimated from the low-energy optical conductivity spectra,
which is proportional to $n/m^*$,
is also small compared with one of the interband transition,
in accordance with such an apparent doped-semiconductor-like behavior. \cite{TVOopt}
Nevertheless, the observed $\gamma$ values are comparable to or even larger than
those of the previously reported $d$-electron heavy-mass systems close to the canonical Mott transition. \cite{TiMIT,VMIT}


\begin{figure}
\begin{center}
\includegraphics*[width=8.5cm]{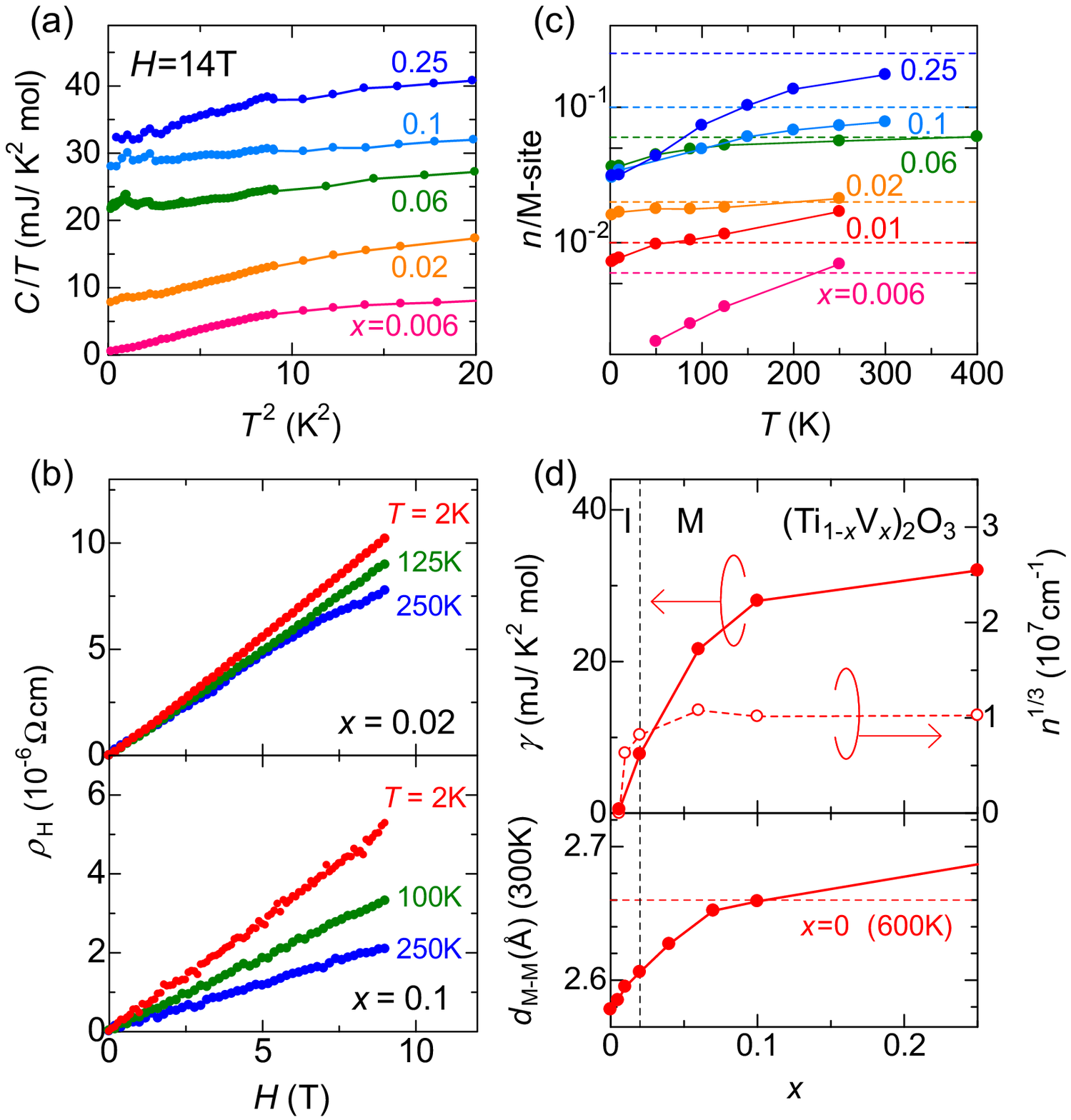}
\caption{(Color online).
(a) Specific heat at $H=14$ T of {\TiVO} plotted in a form of $C/T$ vs. $T^2$ plot.
(b) Magnetic field dependence of the Hall resistivity for {\TiVO} with $x=0.02$ and $x=0.1$ at various temperatures.
(c) Temperature dependence of the carrier density $n$ per one metal (M)-site determined by the Hall coefficient for {\TiVO}.
Each dashed line shows the value of $n=x$.
(d) Doping variation of the electronic specific-heat coefficient $\gamma$
and the cubic root of the carrier density $n^{1/3}$ at 2 K (upper panel).
Their ratio is proportional to the effective mass $m^*$ in the three dimensional system,
and its value in the high-doping metallic region is one or two orders of magnitude larger than that of usual doped semiconductors.
The lower panel shows the doping dependence of the M-M dimer length $d_{\mathrm{M-M}}$ at room temperature. \cite{crysV1}
A horizontal dashed line represents the value estimated for $x=0$ at 600 K. \cite{crysT}
}
\label{fig3}
\end{center}
\end{figure}

In the upper panel of Fig. 3(d), we show the doping variation of
the electronic specific-heat coefficient $\gamma$ and the cubic root of the carrier density $n^{1/3}$ at 2 K.
With lowering temperature, $n$ decreases due to the self trapping of the carriers or the partial valence change of V ions,
as clearly seen in Figs. 3(b) and (c). 
Thus, we show the mobile carrier density $n$ at 2 K rather than the nominal value $x$, for our estimate of the effective mass value.
The ratio between $\gamma$ and $n^{1/3}$ is proportional to the effective mass $m^*$ in the three dimensional system.
In the present case, the rapidly increasing ratio suggests a heavy effective mass in the high-doping metallic region,
e.g., $m^* \sim 75m_0$ at $x=0.1$,
provided that the band of the spin-polarized doped holes is an isotropic parabolic one ($m_0$ being the mass of a free-electron).
The effective mass thus derived should be viewed as semiquantitative,
but is obviously 1 or 2 orders of magnitude larger than that of usual doped semiconductors.
It is dramatically different from the effective mass $m^* \sim 4 m_0$ estimated
by assuming that the low-energy optical conductivity spectra up to 0.2 eV is dominated only by a single Drude component
with constant mass and scattering rate. \cite{TVOopt}
The assumption of the optical effective mass is violated if the carriers are coupled to a low-energy bosonic mode
and show a steep energy variation of the scattering rate and the effective mass. \cite{optmass1,optmass2}
Therefore, the interaction on a lower energy scale ($<$ 0.2 eV) must be the origin of the strongly renormalized thermal effective mass.
Using a relation in the isotropic parabolic band with the fully spin-polarized carriers,
we obtain an approximate value for the Fermi energy $E_{\mathrm{F}} \sim 8$ meV,
which is strongly reduced by the enhanced effective mass as well as the low carrier density.
The magnitude of $E_{\mathrm{F}}$ is comparable to the exchange energy or the energy of
the RKKY interaction ($k_{\mathrm{B}} \theta _{\mathrm{w}} \simeq 2$ meV) and the Zeeman splitting ($2gS_{\mathrm{eff}}\mu _{\mathrm{B}} H \simeq 2$ meV at $H=14$ T). 
These features result in the highly spin-polarized state as readily stabilized by a high (e.g. 14 T) magnetic field.

In general $d$-electron systems, there are some possible origins of the mass renormalization.
Since the local spins are not screened even at the lowest temperature but coupled via the RKKY interaction in {\TiVO},
it is unlikely that the effective mass is enhanced through the Kondo effect.
{\TiVO} is not categorized into the typical Mott transition systems
and the mass enhancement does not occur at the boundary of the insulator-metal transition,
as clearly shown in the doping variation of the ratio between $\gamma$ and $n^{1/3}$ in the upper panel of Fig. 3(d).
This is in contrast to the case of the canonical Mott transition systems, \cite{TiMIT,VMIT}
and thus the strong Coulomb interaction or resultant antiferromagnetic spin correlation
is not likely to be the direct cause of the mass enhancement.
One possible origin is a polaron effect which originates from the strong electron-phonon coupling on the dimer sites
characteristic of the VBS.
In the lower panel of Fig. 3(d), we plot the doping dependence
of the M-M dimer length $d_{\mathrm{M-M}}$ at room temperature (cited from the literature \cite{crysV1}),
which increases monotonically with increasing the V-doping level.
Above $x=0.1$, the dimer length exceeds the value estimated for the thermally induced metallic state at $T=600$ K in $x=0$, \cite{crysT}
where the local singlet pairs are thermally agitated.
As can be seen, the effective mass is greatly enhanced in the high-doping metallic region where the dimerization is weakened.
This suggests the possibility that the carriers in the doped VBS interact most strongly with the dimeric fluctuation or resultant softened phonon in the metallic phase.
If two electrons occupy a dimer, they tend to form a singlet which enhances the dimeric distortion.
If not, the distortion becomes weaker.
In that way, the holes can be strongly coupled with the dimeric distortion, which should lead to the fairly large effective mass.
This mechanism is also consistent with the mass enhancement in the low-energy region of less than 0.2 eV.
As noted above, the large effective mass seems to conversely favor the strong ferromagnetic correlation,
and thus {\TiVO} can be viewed as a magnetic semiconductor,
as materialized by the strong polaronic effect in the doped VBS state.

In summary, we have investigated the heavy-mass electron state in the metallic phase of {\TiVO}.
In this system, the itinerant holes mediate the RKKY interaction between the localized magnetic moments and
become highly spin-polarized at high magnetic fields by coupling with the local spins.
This is also manifested by the large negative magnetoresistance.
Although the carriers are readily ferromagnetically polarized in a high magnetic field (e.g. 14 T),
their effective mass is 1 or 2 orders of magnitude larger than that of conventional doped semiconductors.
One possible mechanism of such mass enhancement is the polaron effect reflecting the strong electron-phonon interaction on the dimer sites.
The doping dependence of the effective mass indicates that the softening of the dimerization is essential for the enhanced interaction leading to the strong mass renormalization.
Conversely, the heavy holes at low density make the highly spin-polarized state more stable at high magnetic fields.
Thus, the present result suggests a new route to the magnetic semiconductor based on the VBS.

We thank N. Nagaosa, N. Furukawa, H. Sakai, and J. Fujioka for fruitful discussions.
This work was partially supported by Grants-in-Aid for Scientific Research (Grants No. 20046004 and No. 20340086) from the MEXT of Japan and JSPS
and Funding Program for World-Leading Innovative R \& D on Science and Technology (FIRST Program). 
M. U. acknowledges support by a Grant-in-Aid for the JSPS Fellowship program (Grant No. 21-5941).

\end{document}